\documentclass[aps,twocolumn,showkeys]{revtex4}

\pdfoutput=1
                             
\usepackage{graphicx}                                       
\usepackage{amssymb}
\usepackage{amsmath}
\usepackage{wasysym}
\usepackage{caption}
\usepackage{multirow}
\usepackage{lipsum}
\usepackage[a4paper,top=15mm,headheight=50mm,%
   bottom=15mm,left=12mm,right=12mm]{geometry}

\newcommand{\eqt}[2]{
    \begin{equation}
             {#2}
             \label{#1}
    \end{equation}
}

\def\ds{\displaystyle}
\def\hs{\vskip 4mm\noindent\  }

\begin{document}

%
%
%
\title{2-Terminal Coaxial Capacitance Measurements and Cable Corrections}
\author{
  F. A. Silveira${}^{}$\footnote{Email address: {\it fsilveira@inmetro.gov.br}} 
}

\affiliation{
  Instituto Nacional de Metrologia, Qualidade e Tecnologia -- Inmetro,\\
  Avenida N. S. das Gra\c cas 50, 25250-020 D. Caxias RJ, Brazil}

\date{\today}

\begin{abstract}

Coaxial 2-terminal bridge comparison is a widely used metrological method in 
the dissemination of value of capacitance up to 1000 pF and $10^4$ rad/s. 
In this range, capacitance can be measured with 
uncertainties as small as few parts in $10^{7}$; however, such measuments must take into account
the corrections due to the impedance of cables, connectors
and ports, which can tipically amount up to $10^{-8}$ $\mu$F/F $-$ and maybe larger.
In this work, we report on how correction factors in 2-terminal pair coaxial bridge capacitance 
measurements are presently measured and calculated at Inmetro.

\end{abstract}

\keywords{
Capacitance standards, coaxial bridges measurements.
}

\maketitle

%
%
%
\section{Introduction}
\label{intro}

The Laboratory of Electrical Standards at Inmetro (Lampe) guarantees the rastreability of
the capacitance unit between 1 and 1000 pF, at frequencies from 400 to $10^4$ rad/s,
through a chain of comparisons of capacitors in a 2-terminal pair (2T) coaxial
bridge \cite{kibble,gregory}. This system is very stable, and went through but small
modifications of its original project ever since it started operating in 2005 \cite{gregory}. 
With this bridge, Lampe compares capacitors with uncertainties smaller than a few parts 
in $10^7$. In this range, corrections due to cabling are not always negligible, and must be 
considered.

\begin{minipage}{\linewidth}
  \makebox[\linewidth]{ 
     \includegraphics[width=.70\linewidth]{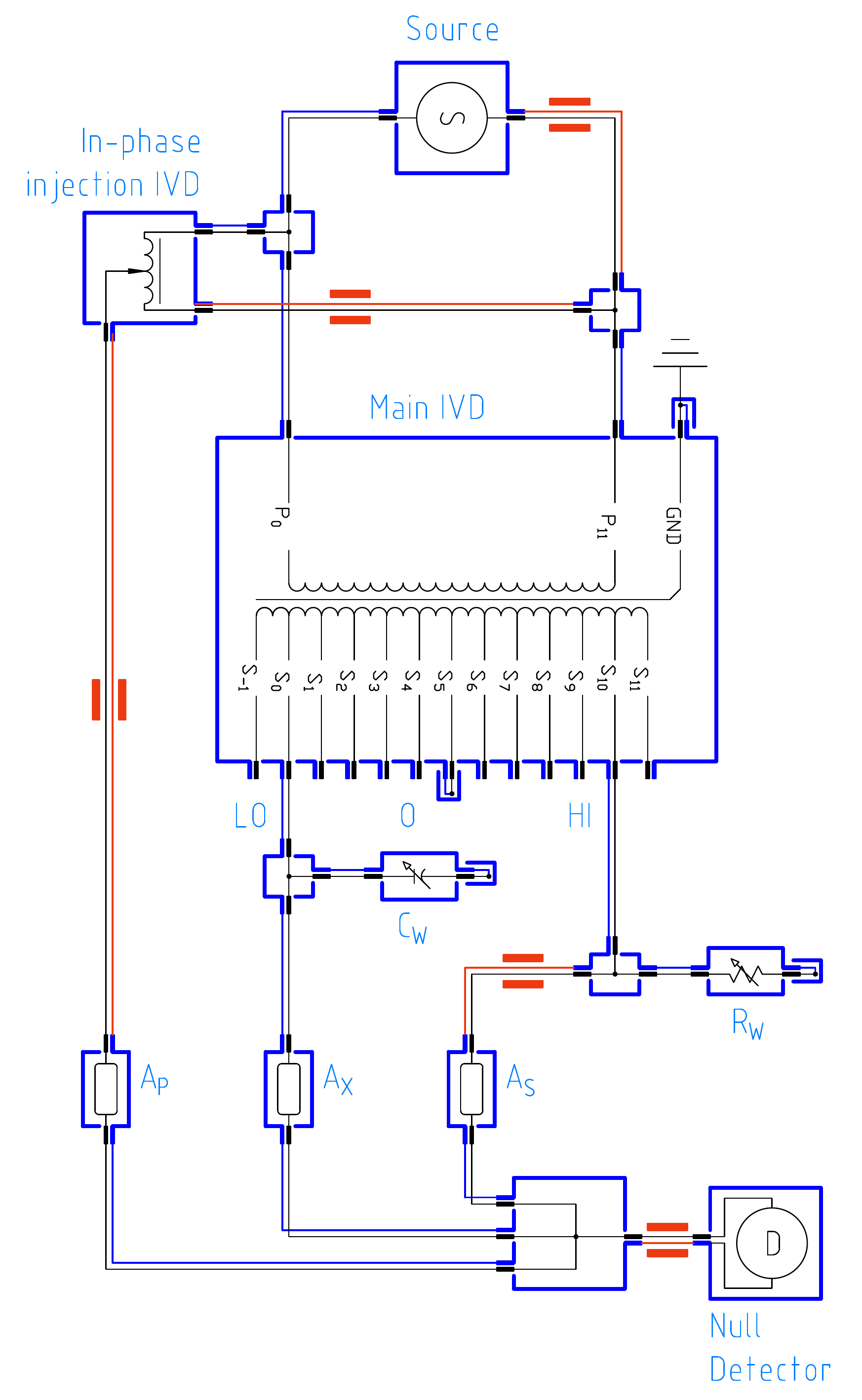}}
  \captionof{figure}{2T coaxial impedance bridge at 1:1 setting.}
  \label{fig1}
\end{minipage}

%

In this context,
correction assessment means determining the relation between the {\bf real} and 
{\bf apparent} impedances of the object in some specific measurement setup. 
The apparent impedance of a standard is expected to be a combination of its real impedance value 
and the shunt admittances of its ports and the cables that couple the standard to the measurement system.
In our case, we seek to 
find the corrections to the capacitance measurements of an object capacitor that's compared
to a known standard capacitor in a coaxial 2T impedance bridge of the type shown in 
Fig. \ref{fig1}.

In this Figure, bold lines represent the outer conductors of the coaxial cables 
and the chassis of the circuit components; the thin lines represent the coax inner 
conductors.

Figure \ref{fig1} illustrates 
the coaxial nature of the circuit and its main inductive voltage divider (IVD), 
the standard and object admittances $A_S$ and $A_X$; 
the auxiliary ground level (Wagner) balance arms, $C_W$ and $R_W$ \cite{wagner};
and the in-phase injection IVD and injection admittance $A_P$.
The five double-bars that appear next to some of the cables represent coaxial chokes, and are meant to 
equalise the inner and outer currents of the choked cables \cite{homan}.

In 2T bridge measurements, a fixed ratio of the main IVD is chosen, usually appropriate
to decadic comparisons of the type 1:1, 1:10 or 1:100 between the capacitances.
Then the injection IVD is adjusted until balance is attained, by null readings
at the detector $D$. This balance is subjected to one main constraint, though: the low potencial 
of the impedances are kept at the ground potencial through the current injected by $C_W$, $R_W$ \cite{wagner}.
The final balance is attained after a few iterated adjustments of $C_W$, $R_W$ 
and the injection IVD ratio \cite{kibble,gregory,wagner}.

At bridge balance, the in-phase circuit admittances in Fig. \ref{fig1} relate through the equation
\eqt{balance}{%
         C'_X= \frac{1-r}{r} C'_S + \frac{\alpha C_P}{r}
}
where $C'_{S,X}$ are the apparent capacitances (related through corrections to the real value of the capacitances 
$C_{S,X}$) being compared, $C_P$ is the capacitance component
of $A_P$, $\alpha$ is the ratio of the in-phase injection IVD \cite{gregory}, and $r$ is the ratio of the main IVD.
Common $r$ settings at the measurement chain in Lampe are $r=1/11$ (1:10 comparisons) and
$r=1/2$ (1:1 comparisons, shown in Fig. \ref{fig1}).

It worths adding that, in practice, there must be a second injection system, in parallel to $A_P$ and the in-phase 
injection IVD, that's not shown in Fig \ref{fig1}.
That system, usually called quadrature injection, has an injection admittance that's 
predominantly conductive, and is meant to balance the small dissipative 
components of the standards \cite{kibble,gregory}.

This work is organized as follows. In Sec. \ref{model} we present the general lumped parameter
model for 2T measurements of a combined impedance formed by the standard and its cables and ports, and 
derive a way of measuring corrections.
in Sec. \ref{corrections} we explore this model in a form suitable to treating capacitance 
2T measurements, and in Sec \ref{measurements} we present some measurement results.
Finally, we comment on these results and draw some conclusions at Sec. \ref{conclusion}.

%
%
%

\section{Equivalent Circuit Model}
\label{model}

Figure \ref{fig2} shows a convenient model in ac 
practice to represent the real impedance $Z$,
and the equivalent admittance of cables and connectors that link it to the measurement circuit. 
In this model, the standard impedance $Z$ is represented with its high and low coaxial ports at B and C, and their respective
shunt admittances $Y_B$ and $Y_C$. Cables connecting these ports to the external circuit, and their
equivalent immittances $Z_1$, $Y_1$ and $Z_2$, $Y_2$, are represented by the meshes at A and D.

\noindent%
\begin{minipage}{\linewidth}
  \makebox[\linewidth]{ 
     \includegraphics[width=.99\linewidth]{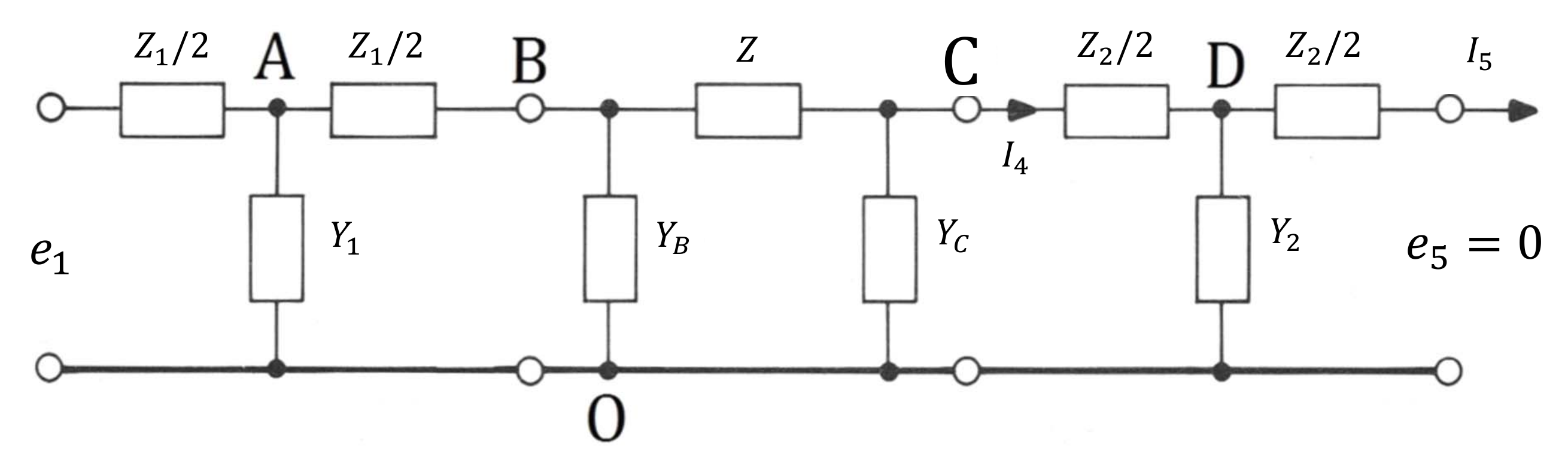}}
  \captionof{figure}{Model of 2T measurement of impedance $Z$.}
  \label{fig2}
\end{minipage}

We associate to each independent loop a clockwise loop current, numbered from $I_1$ to $I_5$ 
(only $I_4$ and $I_5$ are shown in Fig. \ref{fig2}) and we apply the Kirchoff laws 
to find the relation between the real impedance $Z$ and the apparent 
impedance $Z'$. The latter is defined by $Z'=e_1/I_5$, and under the Wagner constraint ($e_5=0$)
the loop equations read

\eqt{loop1}{%
         I_1\left(\frac{Z_1}{2}+\frac{1}{Y_1}\right)-\frac{I_2}{Y_1}-Z'I_5=0
}
\eqt{loop2}{%
         \frac{I_1}{Y_1}-I_2\left(\frac{Z_1}{2}+\frac{1}{Y_1}+\frac{1}{Y_B}\right)+\frac{I_3}{Y_B}=0
}
\eqt{loop3}{%
         \frac{I_2}{Y_B}-I_3\left(Z+\frac{1}{Y_B}+\frac{1}{Y_C}\right)+\frac{I_4}{Y_C}=0
}
\eqt{loop4}{%
         \frac{I_3}{Y_C}-I_4\left(\frac{Z_2}{2}+\frac{1}{Y_C}+\frac{1}{Y_2}\right)+\frac{I_5}{Y_2}=0
}
\eqt{loop5}{%
         \frac{I_4}{Y_2}-I_5\left(\frac{Z_2}{2}+\frac{1}{Y_2}\right)=0
}

By the time of the elaboration of this report, measurements performed 
with a digital RLC bridge on the coax cables used in 2T bridge calibration 
services, and on one of our 100 pF standards, resulted (within combined standard deviation and 
instrument resolution) $L_1=(0.2836\pm 0.0003) \mu$H, $C_1=(72.19\pm 0.05)$ pF
and $C_e=(96.588\pm 0.004)$ pF at $10^4$ rad/s. 
Then we may safely aproximate $Z_{1,2}\sim 10^{-3} \Omega$ and $Y_{B,C}\approx Y_{1,2}\lesssim 10^{-6} S$, 
and contributions to $Z'$ of powers higher than the product of these two quantities 
are expected to be or order $10^{-9}$ or smaller, and are considered negligible.

Solving the simultaneous Eqs. \ref{loop1} to \ref{loop5} is but a tedious calculation, which
can be carried out by a number of methods.
After collecting to the lower order, the first seven terms of the solution to this system are
\begin{equation}
   \begin{aligned}
         Z'=Z\left[ 1 +\frac{Z_1Y_1}{2}\right. &+ Z_1Y_B + \frac{Z_1}{Z} \\
                             &+\frac{Z_2Y_2}{2} + Z_2Y_C + \left.\frac{Z_2}{Z}\right]
         \label{solution}
   \end{aligned}
\end{equation}

As we can expect from the ideal limit $Y_{1,2,B,C}\to 0$ of Fig. \ref{fig2}, $Z_{1,2}$ 
add directly to $Z$, and can't be disregarded unless $Z$ is much larger than $Z_{1,2}$.
Products like $Z_1Y_1$ or $Z_2Y_C$ come from less obvious series associations of
$Z_1/2$ and $Y_{1,B}$, and of $Z_2/2$ and $Y_{2,C}$, that act as loaded voltage
dividers cascaded along the current path. And although such terms can be $10^6$ times smaller
than terms like $Z_{1,2}/Z$, they certainly can't always be considered negligible when 
uncertainties are smaller than parts in $10^7$, as is the case.

This solution differs from the results obtained through a heuristic method applied over the
same model in \cite{kibble} (that ommits the last two terms in Eq. \ref{solution}), 
and from the results in \cite{gregory}, that uses a less general, simpler model.

%
%
%

\section{Corrections to Capacitance Measurements}
\label{corrections}

In the model of Fig. \ref{fig2}, the cable immittances have the general form
$Z_{1,2}=R_{1,2}+j\omega L_{1,2}$ (lossy inductances) and $Y_{1,2,B,C}=j\omega C_{1,2,B,C}$ 
(pure capacitances), where $R_{1,2}$, $L_{1,2}$ and $C_{1,2}$ stand for the series short-circuit 
resistances and inductances, and the parallel open-circuit capacitances of the high and low cables; 
and $C_{B,C}$ are the open-circuit capacitances of the high and low ports of the standards,
at the work angular frequency $\omega$.
As we limit ourselves to the measurement of standard capacitors, we replace $Z^{-1}, Z'^{-1}$
by the pure capacitances $j\omega C$ and $j\omega C'$, and, by operating these
substitutions on Eq. \ref{solution}, we obtain
\begin{equation}
   \begin{aligned}
        \frac{1}{C'}=\frac{1}{C}\biggr[ 1 -&\frac{\omega^2}{2}\left(L_1C_1+L_2C_2\right)
                                                  -{\omega^2}\left(L_1C_B+L_2C_C\right) \\  
                                         -&{\omega^2}\left(L_1+L_2\right)C
                                                   +\frac{j\omega}{2}\left(R_1C_1+R_2C_2\right) \\
                                         +&{j\omega}\left(R_1C_B+R_2C_C\right)
                                                   +{j\omega}\left(R_1+R_2\right)C \biggr]
         \label{solution2}
   \end{aligned}
\end{equation}

As long as, for the purpose of correction calculations, $Z^{-1},Z'^{-1}$ are considered 
pure susceptances from the start, the imaginary component of this equation has 
no physical meaning, as it would imply a nonexistent constraint on the values of
$C$, $R_{1,2}$ and $C_{1,2,B,C}$.
Solving the real part of this equation for the real capacitance $C$, we obtain
\eqt{solution3}{
      C=C'\frac{1-\omega^2\left[\ds\frac{1}{2}\left(L_1C_1+L_2C_2\right)
				-\left(L_1C_B+L_2C_C\right)
			\right]}
				{1+\omega^2 C'\left(L_1+L_2\right)}
}
In Lampe, our main capacitance standards are of gas-dieletric General Radio Co.
models 1404A, 1404B and 1404C (10 to 1000 pF); and fused silica Andeen-Hagerling model 
AH11A (1 and 100 pF). Tipically, $C_{B,C}\sim 100$ pF for these
standards; and $L_{1,2}\sim 0.3$ $\mu$H and $C_{1,2}\sim 100$ pF for a 
metre of good quality, commercially available RG-58 coaxial cable, as results in
the Section \ref{model} show.

With these figures, at $10^4$ rad/s, therefore, ${\omega^2}C'\left(L_1+L_2\right)$ ranges 
from $10^{-10}$ to $10^{-7}$ for capacitances with nominal value between 1 and 1000 pF, 
and we may aproximate Eq. \ref{solution3}, up to first order, to
\eqt{solution4}{
            C=C'\left(1-\Delta\right)
}
where 
\begin{equation}
   \begin{aligned}
        \Delta=\omega^2 \biggr[\frac{1}{2}\left(L_1C_1+L_2C_2\right)
					+&\left(L_1C_B+L_2C_C\right)  \\
					+&\left(L_1+L_2\right)C'
					\biggr]
         \label{delta}
   \end{aligned}
\end{equation}

This makes sense, as we expect $C'>C$, that is, shunt admittances of cables and input ports 
add up to the real capacitance. 
$\Delta$ is the correction to capacitance, and, at $10^4$ rad/s, it is expected to range from $10^{-9}$ 
to $10^{-7}$ for capacitances with nominal value between 1 and 1000 pF.

%
%
%

\section{Measurement and Application of Corrections}
\label{measurements}

We must determine corrections {\bf before} carrying the actual 2T bridge comparison.
To do this, according to Eq. \ref{delta}, for each capacitor standard, we must carry an independent 
set of {\bf seven} different measurements, over the high and low cables, and over the standard 
and its high and low ports, namely: 
   1- the capacitance $C$ of the standard;
   2- the open-circuit capacitance $C_1$ of the high cable; 
   3- the open-circuit capacitance $C_2$ of the low cable; 
   4- the shunt capacitance $C_B$ of the high port of the standard;
   5- the shunt capacitance $C_C$ of the low port of the standard;
   6- the short-circuit inductance $L_1$ of the high cable;
   and finally 7- the short-circuit inductance $L_2$ of the low cable.

\hs%
\begin{minipage}{\linewidth}
  \makebox[\linewidth]{ 
     \includegraphics[width=.99\linewidth]{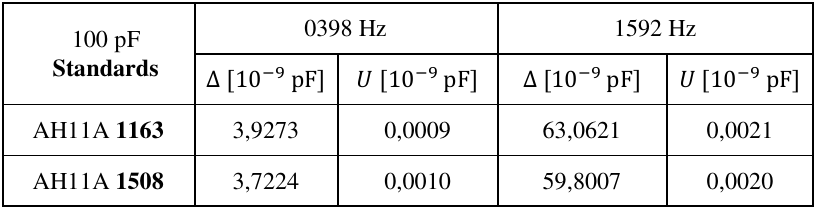}}
  \captionof{table}{Values of $\Delta_X$ for two different units of 100 pF, used as 
					standards in the calibration of 1000 pF units, 
					at frequencies between 2500 and $10^4$ rad/s.}
  \label{tab1}
\end{minipage}
\hs
\begin{minipage}{\linewidth}
  \makebox[\linewidth]{ 
     \includegraphics[width=.99\linewidth]{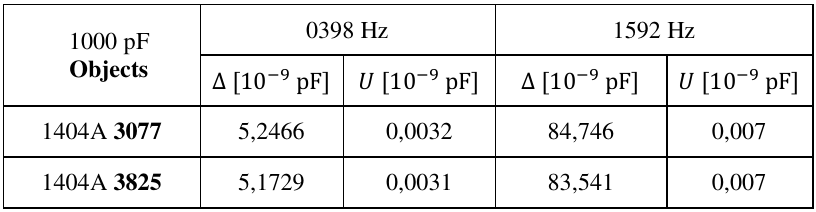}}
  \captionof{table}{Values of $\Delta_X$ for two different units of 1000 pF,
					at frequencies between 2500 and $10^4$ rad/s.}
  \label{tab2}
\end{minipage}
\hs

These measurements can be carried to fairly good uncertainties with a digital bridge like 
the one described in \cite{rlcbridge}, or any other device suitable for some given target uncertainty.
It is also important to mention that $C_{B,C}$ (steps 4 and 5) must be calculated from
intermediary measurements $c_{B,C}$ of the capacitance of the ports B and C, with their complementary port 
(respectively, C and B) short-circuited. Then, using the model of Fig. \ref{fig2}, it's trivial 
to show that $C_{B,C}$ equals the difference between the measured quantities $c_{B,C}-C$.

When we measure $C_X$ in a 2T bridge ($C_S$ as standard), at bridge balance
the apparent capacitances $C'_{S,X}$ relate through Eq. \ref{balance}, so 
we adopt a naming convention as follows. May $\Delta_{S,X}$ be, respectively, the 
corrections to $C_{S,X}$; that is
\eqt{corr1}{
         C_S=C'_S\left(1-\Delta_S\right)
}
\eqt{corr2}{
         C_X=C'_X\left(1-\Delta_X\right)         
}

For each one of the standards, say $C_X$, $\Delta_X$ depends 
on $C'_X$ measured value, 
on the shunt capacitances $C_{B}$ and $C_{C}$ of $C_X$ high and low ports, 
and on the immittances $\omega L_{1,2}$ and $\omega C_{1,2}$ of the high and low cables 
connecting $C_X$ to the system. 
Accordingly, to calculate $\Delta_S$ we must measure $C'_S$, 
the shunt capacitances $C_{B}$ and $C_{C}$ of its ports and the immittances $\omega L_{1,2}$ and
$\omega C_{1,2}$ of its cables.

Once all these quantities are measured, and corrections $\Delta_{S}$ and $\Delta_X$ calculated, 
all at some frequency $\omega$, we substitute Eqs. \ref{corr1} and \ref{corr2} in Eq. \ref{balance}, 
so we get, to first order in $\Delta_{S,X}$,
\eqt{corrbalance}{
                   C_X = \frac{1-r}{r} C_S\left(1+\Delta_S-\Delta_X\right) 
                           + \frac{\alpha C_P}{r}\left(1-\Delta_X\right)
}
where $C_S$ is the known value of the real capacitance of the standard, and $C_X$ is the real
capacitance of the object being compared.

We show in the tables \ref{tab1} and \ref{tab2} the values of $\Delta_{S,X}$ for two pairs of 100 and 
1000 pF standards, used in 1:10 comparisons at the 2T bridge. The units used
as standards in these comparisons are two Andeen-Hagerling AH11A, of nominal value 100 pF,
identified in the tables with their serial numbers, 1163 and 1508. The object units are two
General Radio Co. 1404A capacitors, of nominal value 1000 pF, identified with 
the serials 3077 and 3825.

\

%
%
%

\section{Comments on Results}
\label{conclusion}

The main characteristic Tables \ref{tab1} and \ref{tab2} illustrate is that
corrections don't significantly vary between standards of the same model 
and nominal value. This was an expected result for commercial, serially manufactured 
units, but it is reassuring to quantify it.
In practice, then, we can estimate average values for $C_{B,C}$ between 
units that share same model and nominal value, with negligible loss of information,
and add the dispersion of their values to the final uncertainties of these
parameters.

Tables \ref{tab1} and \ref{tab2} also illustrate the very week deviation
of $\Delta$ of the $\omega^2$ dependence predicted in Eq. \ref{delta}, which reaffirms 
the consistency of the model in Fig. \ref{fig2}, with but small depedence of the 
model parameters on $\omega$.


%
%
%

\onecolumngrid\ \vfill\twocolumngrid

\end{document}